\documentstyle[aps]{revtex}
\begin{document}
\draft
% for two column  activate the line below...
%\twocolumn[\hsize\textwidth\columnwidth\hsize\csname@twocolumnfalse\endcsname
\title
{Secure Communication using Compound Signal from Generalized
Synchronizable Chaotic Systems} 
\author{K.Murali and M. Lakshmanan }
\address{
Centre for Nonlinear Dynamics, Department of Physics, Bharathidasan University,
Tiruchirappalli - 620 024, India}
%\date{May 30, 1997}
\maketitle

\begin{abstract}
By considering generalized synchronizable chaotic systems, the drive-auxiliary
system variables are combined suitably using encryption key functions to 
obtain a compound chaotic signal.  An appropriate feedback loop is constructed in the response-auxiliary
system to achieve synchronization among the variables of the drive-auxiliary and
response-auxiliary systems.  We apply this approach to transmit analog and digital
information signals
in which the quality of the recovered signal is higher and the encoding is more secure.
\end{abstract}
\vskip 10pt

%\newpage
%\narrowtext
%\narrowtext
%\vskip1pc]
Several recent studies have shown the possibility of synchronizing chaotic
systems and its usefulness in secure communications[1-14].  
Among the available chaos synchronization schemes, in the first 
method due to Pecora and Carroll[2]
 a stable subsystem of a chaotic system is synchronized with a separate
 chaotic subsystem under suitable conditions.  This method has been further
 extended to cascading chaos synchronization with multiple stable
 subsystems[1-5].  The second method to achieve chaos synchronization is
 due to the approach of one-way coupling, in which two identical chaotic
 systems are synchronized without requiring to construct any stable
 subsystem[7,8].  In both these approaches only one chaotic signal from the drive
 system is utilized to drive the response systems. These are the most frequently 
 studied schemes where the complete system consists of coupled 
 identical subsystems[1-8].  In these cases the synchronization appears as 
 an actual equality of the corresponding variables of the coupled systems as they                                                
 evolve in time.  This type of synchronization is now called the 
 {\it conventional synchronization}({\bf CS})[15-20].  However a more complicated
 situation arises when {\it coupled nonidentical chaotic systems} are
 considered.  This kind of problem has been recently reported by
 Rulkov {\it et al.}[15] in which a generalization of synchronization 
 for unidirectionally coupled systems was proposed, where two systems are
 called synchronized if a (static) functional relation exists between the 
 states of both the systems.  The resulting synchronization is called 
 {\it generalized synchronization}({\bf GS})[15-17,19,20] and a general theory
 for {\bf GS} of unidirectionally coupled systems was developed in [15]. 
 Recently in ref.[18] the concept of synchronization and communication through 
 compound chaotic signal generated using {\it encryption key functions} have been
 reported.  This approach can be further extended by including different
 auxiliary chaotic systems both at the drive and response systems
 so that both {\bf CS} and {\bf GS} are used where all the 
 {\it drive-auxiliary system} variables are combined suitably
 so that a compound chaotic drive signal is so produced to drive the
 {\it response-auxiliary system}.  A feedback loop in the 
 {\it response-auxiliary system} is constructed
 appropriately to achieve synchronization among the variables of the
 drive and response systems. The importance of the present method is 
 that both the drive and response systems are totally unaltered(unlike the 
 standard available method for {\bf GS}) and the arrangement
 can be easily implemented in practical situations for communications.  Also
 the transmitted compound chaotic signal looks more complex depending upon the
 type of {\it encryption key function} used, thereby improving the security as well.

 The present suggested method of chaos synchronization can be described as follows.
Let us consider the following form of the drive-auxiliary system equations:\\
{\bf drive:}
\begin{equation}
\dot {\bf x}={\bf f(x)}+\epsilon (v(t)-x_j(t)),
\end{equation}
\begin{equation}
\dot {\bf y}={\bf f(y)}+\epsilon (u(t)-y_j(t)),
\end{equation}
{\it auxiliary system:}
\begin{equation}
\dot {\bf z}={\bf g(z)}+\epsilon (y_j(t)-z_j(t)),
\end{equation}
where ${\bf x}$ and ${\bf y}$ are identical chaotic systems and ${\bf z}$ is a
different chaotic system(so called auxiliary system) unidirectionally driven 
by the chaotic
variable $y_j(t)$. If
${\it v(t)}={\it y_j(t)}$ and ${\it u(t)}={\it x_j(t)}$ then Eqs.(1) and (2)
are two identical chaotic systems mutually coupled together with the
signals at the ${\it j}$th component.  This kind of mutually coupled chaotic
systems have been well studied[9].  Here two self-synchronizable chaotic  
systems (1) and (2) are used for experimental operational convenience, which
then are to be used for secure signal transmission applications.  Due to
the present simple set-up, the information signal can easily be injected
as a voltage source experimentally to the chaotic carrier for secure
communication.  We have infact explicitly shown numerically the 
applicability of this kind of set-up in the following discussions.

      Now the two mutually coupled
self-synchronized systems (1) and (2) are considered together with the
one-way coupled auxiliary system (3) as a single
${\it drive}$ system.  The concept of chaos synchronization through
${\it drive-response}$ formalism is then established by considering the
${\it response}$ system equations as\\
{\bf response:}
\begin{equation}
{\bf \dot y_{r}}={\bf f(y_r)}+\epsilon_r (v_r(t)-(y_r)_j(t)).
\end{equation}
{\it auxiliary system:}
\begin{equation}
{\bf \dot z_{r}}={\bf g(z_r)}+\epsilon_r ((y_r)_j(t)-(z_r)_j(t)).
\end{equation}
Here ${\bf y_r}$ is the copy of ${\bf x}$ or ${\bf y}$ and is driven 
by the signal ${\it v_r(t)=v(t)}=y_j(t)$
through one-way coupling[7,8] and ${\bf z_r}$ is the copy of ${\bf z}$.  Here ${\epsilon_r}$ is the one-way coupling
parameter. 
If the maximal Lyapunov exponent of (4) is negative under the
influence of the chaotic signal ${\it v_r(t)}$ added to the ${\it j}$th
component of the response system vector field ${\bf \dot y_{r}}={\bf f(y_r)}$,
then ${\parallel \bf {y_{r}-y}\parallel}\rightarrow 0$ for ${\it t}\rightarrow 
 \infty$. Also if the maximal Lyapunov exponent of (5) is negative under the
influence of the chaotic signal ${\it (y_r)_j(t)}$ added to the ${\it j}$th
component of the auxiliary system vector field ${\bf \dot z_{r}}={\bf g(z_r)}$,
then ${\parallel \bf {z_{r}-z}\parallel}\rightarrow 0$ for ${\it t}\rightarrow 
 \infty$. 
 Synchronization between systems (1) and (2) occurs if the
dynamical system describing the evolution of the difference
${\bf e}={\bf x-y}$,
\begin{equation}
{\bf \dot e}={\bf \dot x}-{\bf \dot y},
\end{equation}
possesses a stable fixed point at the origin ${\bf e=0}$ and synchronization
between (4) and (2) occurs if
\begin{equation}
{\bf \dot e_y}={(\bf \dot y_r - \dot y)},
\end{equation}
possesses a stable fixed point at the origin ${\bf e_y=0}$. 
Also synchronization between (5) and (3) occurs if 
\begin{equation}
{\bf \dot e_z}={(\bf \dot z_r - \dot z)},
\end{equation}
possesses a stable fixed point at the origin ${\bf e_z=0}$. 
 This can be further proved by using (global) Lyapunov functions[3].
In this kind of setup ${\bf CS}$ occurs between ${\bf x}$ \& ${\bf y}$, 
${\bf y_r}$ \& ${\bf y}$ and ${\bf z_r}$ \& ${\bf z}$.  
However ${\bf GS}$ between ${\bf y} \&  {\bf z}$ occurs since there is the 
${\bf CS}$ between ${\bf z}$ and ${\bf z_r}$[15-17,20].
  However, it is not necessary that only one of the drive variables alone be
used for synchronization with the response system(${\it v_r(t)}={\it y_j(t)}$).
 One can also combine and modify the drive signal appropriately, and then
 undo the transformation at the response system for synchronization[18,19].
 Instead of using one drive signal variable, one can transform the drive-auxiliary system
 variables by appropriate linear or nonlinear combinations(which can be 
 treated as {\bf encryption
 key function}) to produce a compound chaotic signal for use as the
 drive signal for synchronization with the response-auxiliary systems.  A
 suitable feedback loop can be deviced in the response-auxiliary system to achieve
 synchronization among the variables of the drive and response systems.  The
 arrangement is illustrated schematically in Figure 1.

We have used the well-known Chua's circuit[9,10] as the drive system.  We 
have considered the auxilary systems as (i) Murali-Lakshmanan-Chua(MLC) 
circuit[1,21] for the non-autonomous case and (ii) Lorenz system for the autonomous case 
to demonstrate the above scheme of chaos synchronization. The first example 
consists of the 
the model equations for the Chua's circuit and MLC circuit which are represented as\\
{\bf drive:}
\begin{eqnarray}
{\dot x_1} & = & \alpha ((x_2-x_1-g_1(x_1))+\epsilon(v(t)-x_1)),\\
{\dot x_2} & = & x_1-x_2+x_3,\\
{\dot x_3} & = & -\beta x_2,\\
{\dot y_1} & = & \alpha ((y_2-y_1-g_1(y_1))+\epsilon(u(t)-y_1)),\\
{\dot y_2} & = & y_1-y_2+y_3,\\
{\dot y_3} & = & -\beta y_2,
\end{eqnarray}
{\it non-autonomous auxiliary system:}
\begin{eqnarray}
{\dot z_1} & = & z_2-g_2(z_1)+\epsilon_r(y_1-z_1),\\
{\dot z_2} & = & -\sigma z_2-z_1+Fsin(\omega t),
\end{eqnarray}
where $g_1(x)=b_1x+0.5(a_1-b_1)(\vert x+1 \vert - \vert x-1 \vert)$ and 
$g_2(x)=b_2x+0.5(a_2-b_2)(\vert x+1 \vert - \vert x-1 \vert)$
and $ a_1=-1.27, b_1=-0.68, a_2=-1.02, b_2=-0.55 ,\sigma =1.015,F=0.15, \omega = 0.75
$, ${\alpha}$=10.0 and ${\beta}$=14.87.
If ${\it v(t)=y_1}$ and ${\it u(t)=x_1}$ then for
appropriate values of mutual coupling parameter ${\epsilon}$$({\epsilon}$=1.3), the above
system of equations (9-14) self-synchronizes.  After synchronization 
$x_1$=$y_1$,$x_2$=$y_2$, and $x_3$=$y_3$. Let us now choose an
appropriate 
drive encryption key function. \\
{\it drive encryption key:}
\begin{equation}
K_d=h(y_1,y_2,y_3,z_1,z_2),
\end{equation}
so that we can generate a sufficiently complicated compound chaotic
signal to be transmitted.  The form of the {\it encryption key function}
 is entirely within our choice. We may choose the function as $z_1,z_1^2,
 y_2z_1^2,y_2^2z_1,y_2y_3z_1z_2, y_2^3z_1$,... .  Then the {\it encryption 
 key function}
is combined with the signal from the drive to generate the compound drive signal
for transmission.\\
{\it compound drive signal:}
\begin{equation}
d(t)=v(t)+K_d = y_1+h,
\end{equation}
Then the response system equations are\\
{\bf response:}\\
{\it response encryption key:}
\begin{equation}
K_r=h(y_{1r},y_{2r},y_{3r},z_{1r},z_{2r}) {\equiv} h_r,
\end{equation}
{\it  regenerated drive signal:}
\begin{equation}
v_r(t)=d(t)-K_r=d(t)-h_r,
\end{equation}
\begin{eqnarray}
{\dot y_{1r}} & = & \alpha ((y_{2r}-y_{1r}-g_1(y_{1r}))+\epsilon_r (v_r(t)-y_{1r})),\\
{\dot y_{2r}} & = & y_{1r}-y_{2r}+y_{3r},\\
{\dot y_{3r}} & = & -\beta y_{2r}.
\end{eqnarray}
{\it auxiliary system:}
\begin{eqnarray}
{\dot z_{1r}} & = & z_{2r}-g_2(z_{1r})+\epsilon_r (y_{1r}-z_{1r}),\\
{\dot z_{2r}} & = & -\sigma z_{2r}-z_{1r}+Fsin(\omega t).
\end{eqnarray}
      
      In the following we first demonstrate the effectiveness of our model analytically
and numerically for the specific choice $h=z_1$ and $h_r=z_{1r}$.  Then
we point out the applicability for more complicated forms of $h$ numerically.
The difference
system(${\bf e=x-y}$) of Eqs.(9-11) and Eqs.(12-14) is
\begin{eqnarray}
{\dot e_1} & = & \alpha ((e_2-e_1-p_ie_1)-2 \epsilon e_1),\\
{\dot e_2} & = & e_1-e_2+e_3,\\
{\dot e_3} & = & -\beta e_2,
\end{eqnarray}
where $p_i=a_1 $ or $ b_1 $(i=1 or 2) which is determined from $g_1(x)$[9].
It is easy to prove that the temporal derivative of the positive definite 
Lyapunov function
\begin{equation}
E=(\beta /2)e_1^2+(\alpha \beta /2)e_2^2 + (\alpha /2)e_3^2,
\end{equation}
is strictly negative,
\begin{eqnarray}
{\dot E} & = & \beta e_1{\dot e_1}+\alpha \beta e_2{\dot e_2} + \alpha e_3{\dot e_3},\\
         & = & -\alpha \beta(e_1-e_2)^2- \alpha \beta(a_1+2 \epsilon)e_1^2 < 0
\end{eqnarray}
for all $ e_1,e_2,e_3 $ when $ \epsilon > -a_1/2 $.  (Note that $a_1<b_1<0$ and $a_1=-1.27$)[9].\\

Also the difference system of Eqs.(12-16) and Eqs.(21-25) for the
specific choice $h=z_1$ and $h_r=z_{1r}$ is given as
\begin{eqnarray}
{\dot e_{y1}} & = & \alpha ((e_{y2}-e_{y1}-p_ie_{y1})-\epsilon_r(e_{y1}+e_{z1})),\\
{\dot e_{y2}} & = & e_{y1}-e_{y2}+e_{y3},\\
{\dot e_{y3}} & = & -\beta e_{y2},\\
{\dot e_{z1}} & = & e_{z2} - q_ie_{z1} + \epsilon_r(e_{y1}-e_{z1})),\\
{\dot e_{z2}} & = & -\sigma e_{z2}-e_{z1},
\end{eqnarray}
where $q_i=a_2 $ or $ b_2 $(i=1 or 2) which is determined from $g_2(x)$.
Here $ e_{y1}=(y_{1r}-y_1),e_{y2}=(y_{2r}-y_2)$, $ e_{y3}=(y_{3r}-y_3)$,
$ e_{z1}=(z_{1r}-z_1)$ and $ e_{z2}=(z_{2r}-z_2)$.

It is easy to prove that the temporal derivative of the positive definite Lyapunov function
\begin{equation}
E=(\beta /2) e_{y1}^2+( \alpha /2) e_{y3}^2+(\alpha \beta /2)(e_{y2}^2+e_{z1}^2+e_{z2}^2),
\end{equation}
is strictly negative,
\begin{eqnarray}
{\dot E} & = & \beta e_{y1}{\dot e_{y1}}+\alpha \beta e_{y2}{\dot e_{y2}}
 + \alpha e_{y3}{\dot e_{y3}}\nonumber \\
& &  +\alpha \beta e_{z1}{\dot e_{z1}}+\alpha \beta e_{z2}{\dot e_{z2}},\\
& = & -\alpha \beta[(e_{y1}-e_{y2})^2+ (a_1+ \epsilon_r) e_{y1}^2\nonumber \\
&  &   +(a_2+ \epsilon_r)e_{z1}^2 +\sigma e_{z2}^2] < 0, 
\end{eqnarray}
for all $ e_{y1},e_{y2},e_{y3},e_{z1} $ and $ e_{z2} $ when $ \epsilon_r > -a_1 $.  
Further the conditional Lyapunov
exponents of the response system Eqs.(21-23) for the given value of one-way
coupling parameter $\epsilon_r$ can be computed through numerical simulations.
For $\epsilon_r$=1.3, the conditional Lyapunov exponents are $(-0.2019,-0.2017,-12.43)$.
Figures 2(a-d) depict the
phase-portraits in the $(y_2$-$y_1)$-plane, $(z_2$-$z_1)$-plane, $(z_1$-$y_1)$-plane
and $(d(t)$-$y_1)$-plane respectively.  The qualitative shape of the 
attractor in Fig.2(d) depends upon the type of {\it encryption key}.
As the maximal conditional Lyapunov exponent is negative,  
synchronization between drive and response systems is achieved as shown in 
Figs.3(a-b) for $ \epsilon_r$=1.3.  Also the conditional Lyapunov exponents
of the {\it auxiliary systems} (15-16) and (24-25) are calculated as
$(-0.8584, -0.855, 0.0)$.

The above scheme of synchronization can be used to construct transmitter-receiver
systems for encoding and masking information data signals. To send the information
signal $s(t)$ from the transmitter to receiver using the familiar {\it chaos 
signal masking technique}[1,3-6,18], now the signals $v(t)$ and $u(t)$ are modified as
$v(t)=y_1+s(t)$ and $u(t)=x_1+s(t)$ respectively.  The significance of this  
type of encoding the
message signal $s(t)$ is not only that it is added just to a 
chaotic carrier
but it also simultaneously drives the self-synchronizing transmitter dynamical
system.  Such an encoding procedure ensures security and also avoids the
typical distortion errors that occur in almost all previous communication
schemes based on chaos synchronization[10].  By employing this scheme, signal
is recovered at the response system Eqs.(19-25) as\\
{\it recovered signal:}
\begin{equation}
r(t)=v_r(t)-y_{1r}=s(t).
\end{equation}
Figure 3(c) and 3(d) show the numerical simulation results of the 
transmitted compound chaotic signal $ d(t) $ and the recovered signal $({\it sine wave}
, 0.02 \sin 0.5t) $ respectively.

            In order to demostrate the signal transmission applications 
using nonlinear {\it encryption key functions}, in the following 
we have used $K_d = h = y_2z_1$.  Then the {\it compound drive signal}
is represented as $ y_1+y_2z_1$ and the {\it regenerated drive signal}
at the response system equations(21-25) is given as $v_r(t)=d(t)-y_{2r}z_{1r}$.
Figures 4(a-c) depict the phase-portrait in $(d(t)-y_1)$-plane for $ \epsilon_r$=1.3,
the transmitted compound chaotic signal $ d(t) $ and the recovered information
signal ({\it sinewave}, 0.03sin0.2t) respectively.  Also Figures 5(a-c) show
the numerical simulation results of the digital information signal $s(t)$,
the transmitted compound chaotic signal $d(t)$ and the recovered digital 
information signal $r(t)$ for $ \epsilon_r $=1.3 respectively for the choice
of $h=y_2z_1$. 

            In the above analysis we have used the auxiliary system as
a nonautonomous system.  One can as well use an autonomous auxiliary
system.  In the following we choose the auxiliary system as the Lorenz system.
The drive equations are the same
as Eqs.(9-14) and also the response equations (21-23).  Now the autonomous auxiliary
system is considered as the Lorenz system driven by the chaotic signal $y_1$
(from Eqs.(12-14)) and the governing model equations are represented as\\
{\it autonomous auxiliary system:}
\begin{eqnarray}
{\dot z_1} & = & -\sigma(z_1-z_2),\\
{\dot z_2} & = & ry_1-z_2-y_1z_3,\\
{\dot z_3} & = & y_1z_2-bz_3, 
\end{eqnarray}
{\it drive encryption key:}
\begin{equation}
K_d=h(z_1,z_2)=z_1,
\end{equation}
{\it compound drive signal:}
\begin{equation}
d(t)=v(t)+K_d = y_1+z_1,
\end{equation}
Then the response system equations are\\
{\bf response:}\\
{\it response encryption key:}
\begin{equation}
K_r=h(z_{1r},z_{2r})=z_{1r},
\end{equation}
{\it  regenerated drive signal:}
\begin{equation}
v_r(t)=d(t)-K_r=d(t)-z_{1r}.
\end{equation}
{\it auxiliary system:}
\begin{eqnarray}
{\dot z_{1r}} & = & -\sigma(z_{1r}-z_{2r}),\\
{\dot z_{2r}} & = & ry_{1r}-z_{2r}-y_{r1}z_{3r},\\
{\dot z_{3r}} & = & y_{1r}z_{2r}-bz_{3r},
\end{eqnarray}.

Here $y_{1r}$ is the signal generated from Eqs.(21-23).  Figures 5(a-b) show
the phase-portrait in  the $(z_3-z_2)-plane$ and $(d(t)-y_1)$-plane respectively for
$\epsilon_r$=1.3, $\sigma$=10, r=28 and b=2.666.
Further by  using this model (Eqs.(41-50)) for secure signal communications, Figure 6(a-c) 
 depict the numerical simulation results of the difference signal  
 $(z_{1r}-z_1)$,
 transmited compound
chaotic signal $d(t)$ and the recovered analog signal(0.02sin0.5t) 
respectively.  As mentioned earlier one can also use more complicated 
encryption key functions without any difficulty. We also note here that 
recently, Chua's circuits(\underline {without} auxiliary systems) 
represented by Eqs.(9-14,20-23) have been used to demonstrate experimentally
the present scheme of secure communication for analog and digital signals
through compound chaotic signal(generated with suitable encryption key
functions)[22].

      In conclusion, we have presented a procedure of achieving an efficient
synchronization using a compound chaotic signal generated from generalized 
synchronizable chaotic systems.  By considering suitable {\it encryption key
functions} a compound drive chaotic signal is produced
and with appropriate feedback-loop at the receiver, synchronization among
the variables of the drive and response has been established.  Also,
its application in secure communications of analog and digital
information signals has
been demonstrated and the information signals have been recovered perfectly.
Due to the present scheme of efficient encoding of message signals with
suitable encryption key functions the compound chaotic signal looks more
complex thereby improving the security of the transmitted signal.

This work has been supported through a research project by the
 Department of Science and Technology, Government of
India.
%\newpage

\newpage
\centerline{FIGURE CAPTIONS}
\begin {figure}
\caption[]{Schematic digram showing the synchronization 
scheme using encryption key function.}
\label{Fig.1}
\end{figure}

\begin {figure}
\caption[]{Generalized synchronization of a Chua's circuit (drive) and
a MLC circuit(auxiliary).  (a) $y_2$ vs $y_1$ of Chua's circuit($\alpha =10.0$,
$\beta =14.87$),
(b) $ z_2$ vs $z_1$ of MLC circuit  
($F=0.15$, $\omega = 0.75$, $\epsilon =
\epsilon_r=1.3$). (c) $z_1$ vs $y_1$,
(d) compound chaotic signal $ d(t)$ vs $y_1$.
Here y1,y2,z1,z2 and d(t)
correspond to $y_1,y_2,z_1,z_2 $ and $d(t)$ respectively in the text.}
\label{Fig.2}
\end{figure}

\begin {figure}
\caption[]{(a) Difference signal $(y_{1r}-y_1)$ vs t for $\epsilon_r =1.3$,
(b) Difference signal $(z_{1r}-z_1)$ vs t,
(c) Transmitted compound chaotic signal $ d(t)=y_1+s(t)+z_1 $.  Here $s(t)=
    0.02sin0.5t$,
(d) Recovered information signal $r(t)$ (using Eq.(40)).
Here y1,y1r,z1,z1r,d(t) and r(t)
correspond to $y_1,y_{1r},z_1,z_{1r},d(t) $ and $r(t)$ respectively in the text.}
\label{Fig.3}
\end{figure}

\begin {figure}
\caption[]{(a) Compound chaotic signal $ d(t)=y_1+s(t)+y_2z_1$ vs $y_1$
$(\alpha =10, \beta =14.87$, $F=0.15$, $\omega = 0.75$, $\epsilon =
\epsilon_r=1.3$).  Here $s(t)=0.03sin0.2t$.
(b) Transmitted compound chaotic signal $ d(t) $,
(c) Recovered information signal
$r(t)$ (using Eq.(40)).}
\label{Fig.4}
\end{figure}

\begin {figure}
\caption[]{(a) Digital information signal,
(b) Transmitted compound chaotic signal $ d(t)=y_1+s(t)+y_2z_1 $,
(c) Recovered digital information signal.}
\label{Fig.5}
\end{figure}

\begin {figure}
\caption[]{(a) $ z_3$ vs $z_2$ of Lorenz equations(41-43)  
($\sigma = 10$, r=28, b=2.666),
(b) compound chaotic signal $ d(t)=(y_1+z_1)$ vs $y_1$.}
\label{Fig.6}
\end{figure}

\begin {figure}
\caption[]{(a) Difference signal $(z_{1r}-z_1)$ vs t,
(b) Transmitted compound chaotic signal $ d(t)=y_1+s(t)+z_1 $.  Here
$s(t)=0.03sin 0.2t$.
(c) Recovered information signal
$r(t)$.}
\label{Fig.7}
\end{figure}
\end{document}